\documentstyle[12pt]{article}
\vspace{1.8cm} \oddsidemargin 0pt \evensidemargin 0pt  \topmargin
-1.5cm \footheight 26pt \footskip 10mm

\begin{document}
\baselineskip 20pt
\begin{center}
\baselineskip=24pt {\Large  Derivation of Nonlinear
Schr\"{o}dinger Equation}

\vspace{1cm} {Xiang-Yao Wu$^{a}$ \footnote{E-mail:wuxy2066@163.com
}, Bai-Jun Zhang$^{a}$, Xiao-Jing Liu$^{a}$, Li-Xiao$^{a}$\\
Yi-Heng Wu$^{a}$, Yan-Wang$^{a}$, Qing-Cai Wang$^{a}$ and Shuang
Cheng$^{b}$
 }
\vskip 10pt \noindent{\footnotesize a. \textit{Institute of
Physics, Jilin Normal University, Siping
136000, China} \\

\footnotesize b. \textit{Department of Physics, Shanghai
University, Shanghai 200444, China}}
\end{center}
\date{}
\renewcommand{\thesection}{Sec. \Roman{section}} \topmargin 10pt
\renewcommand{\thesubsection}{ \arabic{subsection}} \topmargin 10pt
{\vskip 5mm
\begin {minipage}{140mm}
\centerline {\bf Abstract} \vskip 8pt
\par
\indent\\

\hspace{0.3in} We propose some nonlinear Schr\"{o}dinger equations
by adding some higher order terms to the Lagrangian density of
Schr\"{o}dinger field, and obtain the Gross-Pitaevskii (GP)
equation and the logarithmic form equation naturally. In addition,
we prove the coefficient of nonlinear term is very small, i.e.,
the nonlinearity of Schr\"{o}dinger equation is weak.


\end {minipage}

\newpage
\section * {1. Introduction }
\hspace{0.1in} Quantum mechanics is among the most successful of
physical theories. Many of its predictions, especially those of
quantum electrodynamics, have been verified with unparalleled
precision. So it may seem foolhardy to question its most basic
tenet. On the other hand, despite its success, the interpretation
of quantum mechanics remains problematic [1]. There is still no
generally accepted solution to the problem of reduction of the
wave packet. Moreover, there are grave difficulties confronting
relativistic quantum theory, difficulties confronting relativistic
quantum theory, difficulties which have been circumvented but not
eliminated by renormalization theory. There resolution may well
require a thorough reappraisal of the basic principles on which
the theory is founded. The linearity of quantum mechanics,
expressed in the superposition principle is anomalous. Linearity
is a common feature of physical theories, but in all other known
cases it is an approximation. The range over which linearity holds
may be extensive, but is always limited: Maxwell's equations break
down for very intense fields and the linearity of space-time
itself is a weak-field approximation. Practically all physical
phenomena behave nonlinearly when examined over a sufficiently
large range of the dynamical parameters that determine their
evolution.

A number of earlier works that have attempted to extend quantum
theory in a nonlinear way, there are: The work by Broglie-Bohm
nonlinear wave mechanics [2,3] and the Heisenberg spinor unified
field theory [4,5] until present [6-7], the developments of
various nonlinear theories are always remarkable. Thacker
discussed an exact integrability in quantum field theory and
statistical systems, which include the nonlinear Schrodinger model
and equation [6].

A number of different nonrelativistic models of this kind have
been systematically studied by Weinberg, offering also an
assessment of the observational limits on such modifications of
the Schr\"{o}dinger equation [8]. Independently, Doebner and
Goldin and collaborators have also studied nonlinear modifications
of the nonrelativistic Schr\"{o}dinger equation [9]. This was
originally motivated by attempts to incorporate dissipative
effects. Later, however, they have shown that classes of nonlinear
Schr\"{o}dinger equations, including many of those considered
earlier, can be obtained through nonlinear transformations of the
linear quantum mechanical equation. The nonlinear quantum
mechanics has a practical importance in different fields, like
condensed matter, quantum optics and atomic and molecular physics;
even quantum gravity may involve nonlinear quantum mechanics
[10-13]. Another important example is in the modern field of
quantum computing [14-18].

In this paper, we study the nonlinear Schr\"{o}dinger equation by
adding some higher order terms to the Lagrangian density of
Schr\"{o}dinger field, and obtain some nonlinear Schr\"{o}dinger
equation, in which include the famous Gross-Pitaevskii (GP)
equation and the logarithmic form equation.

\section * {2. The Lagrangian of Schr\"{o}dinger field}

\hspace{0.3in}The field Lagrangian $L$ is a functional of field
amplitude $\psi(\vec{r},t)$. It can usually be expressed as the
integral over all space of a Lagrangian density $\cal L$:
\begin{equation}
L=\int{\cal L}(\psi,\nabla\psi,\dot{\psi},t)d^{3}\vec{r}.
\end{equation}
The actual field is derived from Hamilton's variational principle:
\begin{equation}
\delta\int_{t_{1}}^{t_{2}}Ldt=\delta\int_{t_{1}}^{t_{2}}\int{{\cal
L}dt}d^{3}\vec{r}=0,
\end{equation}
where the infinitesimal variation $\delta\psi$ of $\psi$ is
subject to the restrictions
\begin{equation}
\delta\psi(\vec{r},t_{1})=\delta\psi(\vec{r},t_{2})=0.
\end{equation}
From Eqs. (1), (2) and (3), we can obtain the Euler-Lagrange
equation
\begin{equation}
\frac{\partial\cal
L}{\partial\psi}-\frac{\partial}{\partial{t}}(\frac{\partial\cal
L}{\partial\dot{\psi}})-\sum_{i=1}^{3}\frac{\partial}{\partial{x_{i}}}(\frac{\partial{\cal
L}}{\partial(\partial\psi/\partial{x_{i}})})=0.
\end{equation}
If the field Lagrangian density $\cal L$ is given, we can obtain
the classical field equation from  Eq. (4).

From the point of view of second quantization, the Schr\"{o}dinger
equation
\begin{equation}
i\hbar\frac{\partial}{\partial{t}}\psi(\vec{r},t)=-\frac{\hbar^{2}}{2m}\nabla^{2}\psi(\vec{r},t)+V\psi(\vec{r},t),
\end{equation}
is a wave equation of classical field $\psi(\vec{r},t)$, and the
 Schr\"{o}dinger wave field $\psi(\vec{r},t)$ is a kind of complex
 field. The equation of complex field $\psi^{*}(\vec{r},t)$ is
\begin{equation}
-i\hbar\frac{\partial}{\partial{t}}\psi^{*}(\vec{r},t)=-\frac{\hbar^{2}}{2m}\nabla^{2}\psi^{*}(\vec{r},t)+V\psi^{*}(\vec{r},t).
\end{equation}
The Lagrangian density of Schr\"{o}dinger field may be taken to be
\begin{equation}
{\cal
L}=i\hbar\psi^{*}\dot{\psi}-\frac{\hbar^{2}}{2m}\nabla\psi^{*}\cdot\nabla\psi-V\psi^{*}\psi.
\end{equation}
Substituting Eq. (7) into (4), we can obtain the Schr\"{o}dinger
equations (5) and (6).

Since
\begin{equation}
\frac{\partial\cal L}{\partial\psi}=-V\psi^{*} \hspace{0.3in}
\frac{\partial\cal L}{\partial\dot{\psi}}=i\hbar\psi^{*}
\hspace{0.3in} \frac{\partial\cal
L}{\partial(\partial\psi/\partial
x_{i})}=-\frac{\hbar^{2}}{2m}\frac{\partial\psi^{*}}{\partial
x_{i}},
\end{equation}
\begin{equation}
\frac{\partial\cal L}{\partial\psi^{*}}=i\hbar\dot{\psi}-V\psi
\hspace{0.3in} \frac{\partial\cal L}{\partial\dot{\psi^{*}}}=0
\hspace{0.3in} \frac{\partial\cal
L}{\partial(\partial\psi^{*}/\partial
x_{i})}=-\frac{\hbar^{2}}{2m}\frac{\partial\psi}{\partial x_{i}},
\end{equation}
substituting Eqs. (8) and (9) into (4), we have
\begin{equation}
-i\hbar\frac{\partial}{\partial
t}\psi^{*}=-\frac{{\hbar}^{2}}{2m}{\nabla}^{2}\psi^{*}+V\psi^{*},
\end{equation}
\begin{equation}
i\hbar\frac{\partial}{\partial
t}\psi=-\frac{{\hbar}^{2}}{2m}{\nabla}^{2}\psi+V\psi.
\end{equation}
The Eqs. (10) and (11) are Schr\"{o}dinger equations, and then the
Lagrangian density (7) is the Lagrangian density of
Schr\"{o}dinger
 field.
\section * {3.  The nonlinear Schr\"{o}dinger equation from the higher order term }
\hspace{0.3in}The Lagrangian density of
 Schr\"{o}dinger filed $\cal L$ is the function of $\psi$,
$\dot{\psi}$, $\psi^{*}$, $\dot{\psi^{*}}$, $\nabla\psi$ and
$\nabla\psi^{*}$. When it takes the form of Eq. (7), which
includes the terms $\psi^{*}$$\cdot$$\dot{\psi}$,
$\nabla\psi^{*}$$\cdot$$\nabla\psi$ and $\psi^{*}$$\cdot$$\psi$,
it can obtain the linear Schr\"{o}dinger equation (10) or (11).

If the Lagrangian density includes higher order terms, e.g.,
$\nabla\psi^{*}$$\cdot$$\nabla\psi$$\cdot$$\psi^{*}$$\cdot$$\psi$,
$(\nabla\psi^{*})^{2}$$(\nabla\psi)^{2}$, $\ln(\psi^{*}\psi)^{m}$,
$\ln\sqrt[n]{\psi^{*}\psi}$, $(\psi^{*}\psi)^{m}$,
$\sqrt[n]{\psi^{*}\psi}$ and so on, we can obtain the nonlinear
Schr\"{o}dinger equation. In the following, we shall add the
higher order terms to the Lagrangian density  of Schr\"{o}dinger
filed.

(1) adding term
$\nabla\psi^{*}$$\cdot$$\nabla\psi$$\cdot$$\psi$$\cdot$$\psi^{*}$
to the Lagrangian density $\cal L$

When we add term
$\nabla\psi^{*}$$\cdot$$\nabla\psi$$\cdot$$\psi$$\cdot$$\psi^{*}$
to the Lagrangian density, the Lagrangian density  of
Schr\"{o}dinger filed becomes
\begin{equation}
{\cal
L}=i\hbar\psi^{*}\dot{\psi}-\frac{\hbar^{2}}{2m}\nabla\psi^{*}\cdot\nabla\psi-V\psi^{*}\psi+a\nabla\psi^{*}\cdot\nabla\psi\cdot\psi^{*}\psi,
\end{equation}
where $a$ is a constant.

 Since
\begin{equation}
\frac{\partial\cal
L}{\partial\psi}=-V\psi^{*}+a\nabla\psi^{*}\cdot\nabla\psi\cdot\psi^{*},
\end{equation}
\begin{equation}
\frac{\partial\cal L}{\partial\dot{\psi}}=i\hbar\psi^{*},
\end{equation}
\begin{equation}
\frac{\partial\cal
L}{\partial(\partial\psi/\partial{x_{i}})}=-\frac{\hbar^{2}}{2m}\frac{\partial\psi^{*}}{\partial
x_{i}}+a\psi^{*}\psi\frac{\partial\psi^{*}}{\partial x_{i}},
\end{equation}
then
\begin{eqnarray}
\sum_{i=1}^{3}\frac{\partial}{\partial{x_{i}}}(\frac{\partial{\cal
L}}{\partial(\partial\psi/\partial{x_{i}})})&=&-\frac{\hbar^{2}}{2m}\nabla^{2}\psi^{*}
+a\nabla(\psi^{*}\psi)\cdot\nabla\psi^{*}+a\psi^{*}\psi\cdot\nabla^{2}\psi^{*},
\end{eqnarray}
substituting Eqs. (13)-(16) into Eq. (4), we have
\begin{eqnarray}
i\hbar\frac{\partial}{\partial
t}\psi&=&-\frac{\hbar^{2}}{2m}\nabla^{2}\psi+V\psi
-a^{*}\nabla\psi^{*}\cdot\nabla\psi\cdot\psi
\nonumber\\&&+a^{*}\nabla(\psi^{*}\cdot\psi)\cdot\nabla\psi
+a^{*}(\psi^{*}\psi)\nabla^{2}\psi.
\end{eqnarray}
 Eq. (17) is the nonlinear
Schr\"{o}dinger equation adding higher order term
$\nabla\psi^{*}$$\cdot$$\nabla\psi$$\cdot$$\psi$$\cdot$$\psi^{*}$
to the Lagrangian density  of Schr\"{o}dinger filed.

(2) adding term $(\nabla\psi^{*})^{2}$$(\nabla\psi)^{2}$ to the
Lagrangian density $\cal L$

When we add term
$\nabla\psi^{*}$$\cdot$$\nabla\psi$$\cdot$$\psi$$\cdot$$\psi^{*}$
to the Lagrangian density, the Lagrangian density  of
Schr\"{o}dinger field is
\begin{equation}
{\cal
L}=i\hbar\psi^{*}\dot{\psi}-\frac{\hbar^{2}}{2m}\nabla\psi^{*}\cdot\nabla\psi-V\psi^{*}\psi+b(\nabla\psi^{*})^{2}(\nabla\psi)^{2},
\end{equation}
where $b$ is a constant.

Since
\begin{equation}
\frac{\partial\cal L}{\partial\psi}=-V\psi^{*},
\end{equation}
\begin{equation}
\frac{\partial\cal L}{\partial\dot{\psi}}=i\hbar\psi^{*},
\end{equation}
\begin{equation}
\frac{\partial{\cal
L}}{\partial(\partial\psi/\partial{x_{i}})}=-\frac{\hbar^{2}}{2m}\frac{\partial\psi^{*}}{\partial
x_{i}}+2b(\nabla\psi^{*})^{2}(\frac{\partial\psi}{\partial
x_{i}}),
\end{equation}
then
\begin{eqnarray}
\sum_i\underline{}\frac{\partial}{\partial{x_{i}}}(\frac{\partial{\cal
L}}{\partial(\partial\psi/\partial{x_{i}})})&=&-\frac{\hbar^{2}}{2m}\nabla^{2}\psi^{*}
+4b\nabla^{2}\psi^{*}\nabla\psi^{*}\cdot\nabla\psi+2b(\nabla\psi^{*})^{2}\nabla^{2}\psi,
\end{eqnarray}
substituting Eqs. (19)-(22) into Eq. (4), we have
\begin{eqnarray}
i\hbar\frac{\partial}{\partial
t}\psi&=&-\frac{\hbar^{2}}{2m}\nabla^{2}\psi+V\psi
+4b^{*}\nabla^{2}\psi\nabla\psi\cdot\nabla\psi^{*}+2b^{*}(\nabla\psi)^{2}\nabla^{2}\psi^{*}.
\end{eqnarray}
Eq. (23) is the nonlinear Schr\"{o}dinger equation adding higher
order term $(\nabla\psi^{*})^{2}$$(\nabla\psi)^{2}$ to the
Lagrangian density of Schr\"{o}dinger filed.

(3) adding term $\ln(\psi^{*}\psi)^{m}$ to the Lagrangian density
$\cal L$

When we add term $\ln(\psi^{*}\psi)^{m}$ to the Lagrangian
density, the Lagrangian density of Schr\"{o}dinger field becomes
\begin{equation}
{\cal
L}=i\hbar\psi^{*}\dot{\psi}-\frac{\hbar^{2}}{2m}\nabla\psi^{*}\cdot\nabla\psi-V\psi^{*}\psi+c\ln(\psi^{*}\psi)^{m},
\end{equation}
where $c$ is a constant.
 \hspace{0.3in}

 Since
\begin{equation}
\frac{\partial\cal
L}{\partial\psi}=-V\psi^{*}+cm\frac{\psi^{*}}{\psi^{*}\psi},
\end{equation}
\begin{equation}
\frac{\partial\cal L}{\partial\dot{\psi}}=i\hbar\psi^{*},
\end{equation}
\begin{equation}
\frac{\partial{\cal
L}}{\partial(\partial\psi/\partial{x_{i}})}=-\frac{\hbar^{2}}{2m}\frac{\partial\psi^{*}}{\partial
x_{i}},
\end{equation} \hspace{0.3in}
 then
\begin{equation}
\sum_{i}\frac{\partial}{\partial x_{i}}(\frac{\partial\cal
L}{\partial(\partial\psi/\partial
x_{i})})=-\frac{\hbar^{2}}{2m}\nabla^{2}\psi^{*},
\end{equation}
substituting Eqs. (24)-(28) into (4), we have
\begin{equation}
i\hbar\frac{\partial\psi}{\partial
t}=-\frac{\hbar^{2}}{2m}\nabla^{2}\psi+V\psi-c^{*}m\frac{\psi}{\psi^{*}\psi}.
\end{equation}
Eq. (29) is the nonlinear Schr\"{o}dinger equation adding higher
order term $\ln(\psi^{*}\psi)^{m}$ to the Lagrangian density of
Schr\"{o}dinger field.

(4) adding term $\sqrt[n]{\psi^{*}\psi}$ to the Lagrangian density
$\cal L$

When we add term  $\sqrt[n]{\psi^{*}\psi}$ to the Lagrangian
density, the Lagrangian density of Schr\"{o}dinger field becomes
\begin{equation}
{\cal L}=i\hbar\psi^{*}\dot{\psi}
-\frac{\hbar^{2}}{2m}\nabla\psi^{*}\cdot\nabla\psi-V\psi^{*}\psi+d\sqrt[n]{\psi^{*}\psi}.
\end{equation}
 where $d$ is a constant.

 Since
\begin{equation}
\frac{\partial\cal
L}{\partial\psi}=-V\psi^{*}+\frac{1}{n}d(\psi^{*}\psi)^{\frac{1}{n}-1}\cdot\psi^{*},
\end{equation}
\begin{equation}
\frac{\partial\cal L}{\partial\dot{\psi}}=i\hbar\psi^{*},
\end{equation}
\begin{equation}
\frac{\partial{\cal
L}}{\partial(\partial\psi/\partial{x_{i}})}=-\frac{\hbar^{2}}{2m}\frac{\partial\psi^{*}}{\partial
x_{i}},
\end{equation}
then
\begin{equation}
\sum_{i}\frac{\partial}{\partial x_{i}}(\frac{\partial\cal
L}{\partial(\partial\psi/\partial
x_{i})})=-\frac{\hbar^{2}}{2m}\nabla^{2}\psi^{*},
\end{equation}
substituting Eqs. (31)-(34) into (4), we have
\begin{equation}
i\hbar\frac{\partial\psi}{\partial
t}=-\frac{\hbar^{2}}{2m}\nabla^{2}\psi+V\psi-\frac{1}{n}d^{*}(\psi^{*}\psi)^{\frac{1}{n}-1}\cdot\psi.
\end{equation}
Eq. (35) is the nonlinear Schr\"{o}dinger equation adding higher
order term $\sqrt[n]{\psi^{*}\psi}$ to the Lagrangian density of
Schr\"{o}dinger field.

(5) adding term $(\psi^{*}\psi)^{m}$ to the Lagrangian density
$\cal L$

 When we add term $(\psi^{*}\psi)^{m}$ to the Lagrangian
density, the Lagrangian density of Schr\"{o}dinger field becomes
\begin{equation}
{\cal
L}=i\hbar\psi^{*}\dot{\psi}-\frac{\hbar^{2}}{2m}\nabla\psi^{*}\cdot\nabla\psi-V\psi^{*}\psi+e(\psi^{*}\psi)^{m}.
\end{equation}
 where $e$ is a constant.

Since
\begin{equation}
\frac{\partial\cal L}{\partial\psi}=-V\psi^{*}+em(\psi^{*}\psi
)^{m-1}\psi^{*},
\end{equation}
\begin{equation}
\frac{\partial\cal L}{\partial\dot{\psi}}=i\hbar\psi^{*},
\end{equation}
\begin{equation}
\frac{\partial{\cal
L}}{\partial(\partial\psi/\partial{x_{i}})}=-\frac{\hbar^{2}}{2m}\frac{\partial\psi^{*}}{\partial
x_{i}},
\end{equation}
then
\begin{equation}
\sum_{i}\frac{\partial}{\partial x_{i}}(\frac{\partial\cal
L}{\partial(\partial\psi/\partial
x_{i})})=-\frac{\hbar^{2}}{2m}\nabla^{2}\psi^{*},
\end{equation}
substituting Eqs. (37)-(40) into (4), we have
\begin{equation}
i\hbar\frac{\partial\psi}{\partial
t}=-\frac{\hbar^{2}}{2m}\nabla^{2}\psi+V\psi-e^*m|\psi|^{2(m-1)}\psi.
\end{equation}
Eq. (41) is the nonlinear Schr\"{o}dinger equation adding higher
order term $(\psi^{*}\psi)^{m}$ to the Lagrangian density of
Schr\"{o}dinger field.

(6) adding term $\psi\psi^{*}\ln(\psi\psi^{*})$ to the lagrangian
density $\cal L$

When we add term $\psi\psi^{*}\ln(\psi\psi^{*})$ to the Lagrangian
density, the lagrangian density of Schr\"{o}dinger field is
\begin{equation}
{\cal
L}=i\hbar\psi^{*}\dot{\psi}-\frac{\hbar^{2}}{2m}\nabla\psi^{*}\cdot\nabla\psi-V\psi^{*}\psi+f\psi\psi^{*}\ln(\psi\psi^{*}).
\end{equation}
Where $f$ is a constant.\\
Since
\begin{equation}
\frac{\partial\cal
L}{\partial\psi}=-V\psi^{*}+f\psi^{*}\ln(\psi\psi^{*})+f\psi^{*},
\end{equation}
\begin{equation}
\frac{\partial\cal L}{\partial\dot{\psi}}=i\hbar\psi^{*},
\end{equation}
\begin{equation}
\frac{\partial{\cal
L}}{\partial(\partial\psi/\partial{x_{i}})}=-\frac{\hbar^{2}}{2m}\frac{\partial\psi^{*}}{\partial
x_{i}},
\end{equation}
substituting Eqs. (43)-(45) into (4), we have
\begin{equation}
i\hbar\frac{\partial\psi}{\partial
t}=-\frac{\hbar^{2}}{2m}\nabla^{2}\psi+V\psi-f^{*}\ln(|\psi|^{2})\psi+f^{*}\psi.
\end{equation}
Eq. (46) is the nonlinear Schr\"{o}dinger equation adding higher
order term $f\psi\psi^{*}\ln(\psi\psi^{*})$ to the Lagrangian
density of Schr\"{o}dinger field.

(7) adding term $\psi\psi^{*}\ln\frac{\psi}{\psi^{*}}$ to the
Lagrangian density $\cal L$

When we add term $\psi\psi^{*}\ln\frac{\psi}{\psi^{*}}$ to the
Lagrangian density, the Lagrangian density of Schr\"{o}dinger
field becomes
\begin{equation}
{\cal
L}=i\hbar\psi^{*}\dot{\psi}-\frac{\hbar^{2}}{2m}\nabla\psi^{*}\cdot\nabla\psi-V\psi^{*}\psi+g\psi\psi^{*}\ln\frac{\psi}{\psi^{*}}.
\end{equation}
Where $g$ is a constant.\\
Since
\begin{equation}
\frac{\partial\cal
L}{\partial\psi}=-V\psi^{*}+g\psi^{*}\ln\frac{\psi}{\psi^{*}}+g\psi^{*},
\end{equation}
\begin{equation}
\frac{\partial\cal L}{\partial\dot{\psi}}=i\hbar\psi^{*},
\end{equation}
\begin{equation}
\frac{\partial{\cal
L}}{\partial(\partial\psi/\partial{x_{i}})}=-\frac{\hbar^{2}}{2m}\frac{\partial\psi^{*}}{\partial
x_{i}},
\end{equation}
substituting Eqs. (47)-(50) into (4), we have
\begin{equation}
i\hbar\frac{\partial\psi}{\partial
t}=-\frac{\hbar^{2}}{2m}\nabla^{2}\psi+V\psi-g^{*}(\ln\frac{\psi}{\psi^{*}})^{*}\psi-g^{*}\psi.
\end{equation}
Eq. (51) is the nonlinear Schr\"{o}dinger equation adding higher
order term $\psi\psi^{*}\ln\frac{\psi}{\psi^{*}}$ to the
Lagrangian density of Schr\"{o}dinger field.

When $m=2$, the nonlinear equation (41) is the same as the
Gross-Pitaevskii (GP) equation[19]
\begin{equation}
i\hbar\frac{\partial\psi}{\partial
t}=-\frac{\hbar^{2}}{2m}\nabla^{2}\psi+V\psi-g|\psi|^{2}\psi.
\end{equation}
\hspace{0.3in}The nonlinear equation (46) is the same as the
equation taken in the Logarithmic form in Ref. [20]
\begin{equation}
i\hbar\frac{\partial\psi}{\partial
t}=-\frac{\hbar^{2}}{2m}\nabla^{2}\psi+V\psi-b\psi\ln|\psi|^{2}.
\end{equation}
\hspace{0.3in}The nonlinear equation (51) is similar as the
equation in Ref. [19]
\begin{equation}
i\hbar\frac{\partial\psi}{\partial
t}=-\frac{\hbar^{2}}{2m}\nabla^{2}\psi+V\psi-i\hbar\frac{b}{2m}(\ln\frac{\psi}{\psi^{*}})\psi.
\end{equation}
\hspace{0.3in}In the following, we shall discuss the nonlinear
equation (41).

When $m=2$, the equation (41) is
\begin{equation}
i\hbar\frac{\partial\psi}{\partial
t}=-\frac{\hbar^{2}}{2m}\nabla^{2}\psi+V\psi-2e^*|\psi|^{2}\psi=0,
\end{equation}
and the Lagrangian density of Schr\"{o}dinger field is
\begin{equation}
{\cal
L}=i\hbar\psi^{*}\dot{\psi}-\frac{\hbar^{2}}{2m}\nabla\psi^{*}\cdot\nabla\psi-V\psi^{*}\psi+e(\psi^{*}\psi)^{2}.
\end{equation}
\hspace{0.3in}We can determine the constant $e$ by dimension
analysis. The Lagrangian density $\cal L$ of Schr\"{o}dinger field
is the dimension of energy density, i.e., $[E][L]^{-3}$, and
$\psi$ dimension is $[L]^{-\frac{3}{2}}$. In Eq. (56), the
dimension of constant $e$ is $[E][L]^{3}$, and the dimension of
constant $\frac{\hbar^{3}}{m^{2}c}$ is $[E][L]^{3}$. The constant
$e$ can be written as
\begin{equation}
e=g\frac{\hbar^{3}}{m^{2}c}.
\end{equation}
Where the constant $g$ is dimensionless. The Eq. (55) becomes
\begin{equation}
i\hbar\frac{\partial\psi}{\partial
t}=-\frac{\hbar^{2}}{2m}\nabla^{2}\psi+V\psi-g\frac{2\hbar^{3}}{m^{2}c}|\psi|^{2}\psi=0.
\end{equation}
Since $\frac{\hbar^{3}}{m^{2}c}\ll1$, the nonlinearly term is very
small, i.e., the nonlinearity of Schr\"{o}dinger equation (58) is
very weak.

When $m=3$, the equation (41) is
\begin{equation}
i\hbar\frac{\partial\psi}{\partial
t}=-\frac{\hbar^{2}}{2m}\nabla^{2}\psi+V\psi-3e^{*}|\psi|^{4}\psi=0,
\end{equation}
and the Lagrangian density of Schr\"{o}dinger field is
\begin{equation}
{\cal
L}=i\hbar\psi^{*}\dot{\psi}-\frac{\hbar^{2}}{2m}\nabla\psi^{*}\cdot\nabla\psi-V\psi^{*}\psi+e(\psi^{*}\psi)^{3}.
\end{equation}
\hspace{0.3in}The dimension of constant $e$ is $[E][L]^{6}$, and
the dimension of constant $\frac{\hbar^{6}}{m^{5}c^{4}}$ is
$[E][L]^{6}$ also. So, the constant $e$ can be written as
\begin{equation}
e=g'\frac{\hbar^{6}}{m^{5}c^{4}},
\end{equation}
where the constant $g'$ is dimensionless. The Eq. (59) becomes
\begin{equation}
i\hbar\frac{\partial\psi}{\partial
t}=-\frac{\hbar^{2}}{2m}\nabla^{2}\psi+V\psi-g'\frac{3\hbar^{6}}{m^{5}c^{4}}|\psi|^{4}\psi=0.
\end{equation}
Obviously, the nonlinearity of Schr\"{o}dinger equation (62) is
weak also.

\section * {4. Conclusion}
\hspace{0.3in}In this paper, we add higher order terms to the
Lagrangian density of Schr\"{o}dinger field and obtain some
nonlinear Schr\"{o}dinger equations. These equations include
nonlinear Schr\"{o}dinger equations proposed by some authors, such
as Gross-Pitaevskii(GP) equation and the logarithmic form
nonlinear equation. In addition, we prove the coefficient of
nonlinear term is very small, i.e., the nonlinearity of
Schr\"{o}dinger equation is weak. We think these nonlinear
Schr\"{o}dinger equations can be used widely in condensed matter
and nonlinear fields.

\end{document}